\documentclass[prb,amsmath,superscriptaddress,citeautoscript,twocolumn,
showkeys,showpacs,floatfix]{revtex4}

\usepackage{bm}
\usepackage{amssymb}
\usepackage{graphicx}

\begin{document}

\title{Noise spectra of an interacting quantum dot}

\author{N. Gabdank}

\affiliation{Department of Physics, Ben Gurion University, Beer
Sheva 84105, Israel}

\author{E. A. Rothstein}
\email{rotshtei@bgu.ac.il}

\affiliation{Department of Physics, Ben Gurion University, Beer
Sheva 84105, Israel}

\author{O. Entin-Wohlman}

\altaffiliation{Also at Tel Aviv University}

\affiliation{Department of Physics and the Ilse Katz Center for
Meso- and Nano-Scale Science and Technology, Ben Gurion University, Beer
Sheva 84105, Israel}

\author{A. Aharony}

\altaffiliation{Also at Tel Aviv University}

\affiliation{Department of Physics and the Ilse Katz Center for
Meso- and Nano-Scale Science and Technology, Ben Gurion University, Beer
Sheva 84105, Israel}

\date{\today}

\begin{abstract}
We study the noise spectra of a many-level quantum dot coupled to two electron reservoirs, when interactions are taken into account only on the dot within the Hartree-Fock approximation. The dependence of the noise spectra on the interaction strength, the coupling to the leads,  and  the chemical potential is  derived. For zero bias and zero temperature, we find that as a function of the (external) frequency, the noise exhibits  steps and dips at frequencies reflecting the internal structure of the energy levels on the dot.  Modifications due to a finite bias and finite temperatures are investigated for a non-interacting two-level dot. Possible relations to experiments are pointed out.
\end{abstract}

% 73.21.La  Quantum dots
% 72.10.-d  Theory of electronic transport; scattering mechanisms
% 72.70.+m  Noise processes and phenomena

\maketitle

\section{Introduction}

\label{INTRO}

The  fluctuations of the electric current flowing  through a
mesoscopic system,  connected to several electronic
reservoirs (e.g., a constriction between two reservoirs) are
described by the {\it noise spectrum}. This spectrum contains
valuable information on the characteristics of the
 system, which cannot be detected by dc conductance measurements.
 Examples include information on Coulomb interactions, on the quantum statistics of charge carriers,
 on the number of available transport channels and  on electrons entanglement.
 \cite{LANDAUER,IMRY,BEENAKKER,BLANTER} Another example comes from heat transport:
recently it was shown that at finite frequencies, the noise associated with
correlations of energy currents in a mesoscopic constriction  is not related to the thermal conductance of the
constriction by the  fluctuation-dissipation theorem,  but
contains additional information on the electric conductance, especially at very low temperatures
where the heat conductance is vanishingly small. \cite{AVERIN}

The \emph{non-symmetrized} noise spectrum is given by the
Fourier transform of the current-current correlation \cite{IMRY}
\begin{align}
 C^{}_{\alpha \alpha'}(\omega)= \int _{-\infty} ^{\infty} dt e^{i\omega t}\langle
\delta \hat{I}^{}_{\alpha}(t)\delta \hat{I}^{}_{\alpha '}(0) \rangle\ ,
\label{Unsymmetrized noise}
\end{align}
where $\alpha$ and $\alpha '$ mark the leads carrying the current
between the electronic reservoirs and the system at hand. In Eq.
(\ref{Unsymmetrized noise}), $\delta \hat{I}_\alpha \equiv
\hat{I}_{\alpha} - \langle\hat{I}_{\alpha}\rangle$, where
$\hat{I}_{\alpha}$ is the operator of the current emerging from
the  $\alpha$th reservoir, and the average (denoted by
$\langle...\rangle$) is taken over the reservoirs' states.
Mesoscopic devices typically operate at low temperatures $T$,
allowing measurements at relatively high frequencies, which obey
$\hbar\omega>k^{}_BT$. In this range one may observe  quantum
deviations from the more familiar thermal
fluctuations.\cite{CLERK} The non-symmetrized noise
spectrum can be interpreted as due to absorption of energy
(positive frequencies) or emission of energy (negative
frequencies) by the noise source \cite{GAVISH, AGUADO, CLERK}.
This interpretation does not hold for the {\em symmetrized} version
of the noise spectrum, given by $C^{\rm{sym}}_{\alpha
\alpha'}(\omega)= [C^{}_{\alpha
\alpha'}(\omega)+C^{}_{\alpha \alpha'}(-\omega)]/2$.

Measuring the non-symmetrized noise spectrum is a
challenging task. It involves a ``quantum detector", where the
measured frequency matches an energy gap in the detector.
 Deblock \textit{et
al}.\cite{DEBLOCK} used a superconductor-insulator-superconductor
(SIS) junction as a noise detector, and the measured frequency
corresponded to the SIS energy gap. Later on, Gustavsson
\textit{et al}. \cite{GUSTAVSSON} used a two-level system (TLS) as
a detector, where the measured frequency corresponded to the
energy difference between the two levels. A similar TLS system was
used in Ref. ~\onlinecite{DEBLOCK 2} where
the {\em equilibrium} non-symmetrized noise spectrum was
detected. More information about quantum noise measurements can be
found in a recent review by Clerk \textit{et al}. \cite{CLERK}

%%%%%%%%%%% The noise spectrum of a quantum dot
In  previous work we have used the
scattering formalism to show that the noise spectra of a
non-interacting quantum dot exhibit features such as steps and
dips, \cite{STEPS AND DIPS,ME}  whose relative locations reflect the
resonance levels of the quantum dot and  the energy spacings
in-between them. A similar dip structure was found by Dong {\it
et al.} \cite{DONG} for two non-interacting or highly interacting
dots placed on the branches of an Aharonov-Bohm interferometer, using a
quantum master equation. The dip was supplanted by a peak
once the repulsive interaction was strong enough.
These authors also found that
the depth of this dip oscillates with the Aharonov-Bohm
flux \cite{DONG}, confirming the results of Ref. ~\onlinecite{STEPS AND DIPS},
which pointed out that the dip is due to interference between
resonance levels.  Later on, Wu and Timm \cite{WU} used the
Floquet master equation to study the noise of a single-level
quantum dot driven by either an ac gate voltage or by a rotating
magnetic field. In both cases they found a peak or a dip,
depending on the nature of the ac driving field, at a frequency
matching the energy difference between pairs of
Floquet channels (associated with transport channels). Recently, Marcos {\it et al.} \cite{MARCOS}
used a quantum master equation to study  a system similar  to the
one analyzed  in Refs. ~\onlinecite{STEPS AND DIPS} and ~\onlinecite{ME}, and found
an interference-induced dip in the noise spectrum of a quantum dot coupled
symmetrically  to two leads.
This result contradicts that of Ref.  ~\onlinecite{STEPS AND DIPS}, which
 found that such a dip appears only when the leads
 are coupled asymmetrically to the dot. We explore this contradiction in Sec.  \ref{APPENDIX B}, where we confirm that the  dip disappears
for a symmetric coupling of the dot to the leads. Moreover, Marcos {\it et al.} \cite{MARCOS} argue that the dip is immune
to the application of a finite bias voltage and to finite temperatures. In Sec.  \ref{APPENDIX B} we find,
 in agreement with the results of Ref.
~\onlinecite{MARCOS}, that the dip location is unchanged by the application of a finite bias
and$\backslash$or temperature; its depth though is affected, and it may even turn into a peak. Very recently, 
Hammer and Belzig \cite{HAMMER} extended the model of Ref. ~\onlinecite{ME} by considering a periodic ac bias voltage and found a similar step-like structure for the noise as the ones found in Refs. ~\onlinecite{STEPS AND DIPS, ME} and to the one found here.

%%%%%%%%%% The asymmetry parameter
The noise spectrum of a quantum dot, as a function of the frequency, is indeed rather
sensitive to
the coupling  of the dot to the leads. \cite{STEPS AND DIPS,ME, DONG} This sensitivity has been analyzed
by Thielmann {\it et al.} \cite{THIELMANN} and measured by   Gustavsson {\it
et al.} \cite{GUSTAVSSON2} who have monitored the
shot  noise as a function of the asymmetry of the dot-leads
couplings.
This  asymmetry  turned out to be
in particular important in the experiment of K\"{o}nemann \textit{et al.} \cite{KYNEMANN} where the current through a highly asymmetrical quantum dot was measured as a function of the applied bias. These authors attributed the observed dependence  of
the broadening of  the slope of the steps in the $I-V$ characteristics  on the polarity of the  bias voltage
to electronic interactions taking place on the dot: They argue that the bias voltage determines whether the
energy level, located between the two chemical potential of the leads, is predominantly empty or occupied. \cite{KYNEMANN} We expound on this point in the Appendix.

%%%%%%%%%%% The Hartree Fock approximation

In the present paper we study the noise spectrum of the current passing through an
interacting spinless quantum dot, coupled to two electronic
reservoirs via two leads, and thus
extend our previous calculations to include the Coulomb interaction.
That interaction is treated within the Hartree-Fock (HF) self-consistent
approximation; by this, the  Hamiltonian  is turned effectively into a single-particle one,
enabling us to employ the scattering formalism in our calculations.
Those are augmented by a self-consistent determination
of the Hamiltonian effective energies and  hopping terms among the levels. Having mapped the Hamiltonian to an effective single-particle Hamiltonian we have constricted the effect of the Coulomb interaction. For example, the Kondo effect, a many-body phenomenon associated with the screening of the spin of the quantum dot (for a dot with an odd number of electrons) by the conduction electrons of the leads at low temperatures, can not be observed in our formalism. Zarchin \textit{et al.} \cite{ZARCHIN} have measured the low-frequency noise spectrum to extract the backscattered charge from a two terminal system with a quantum dot in the Kondo regime and found it to be highly non trivial. Recently, using a real-time functional renormalization group approach, Moca \textit{et al}. \cite{MOCA} found two dips in the noise spectrum that originate from the Kondo effect.

Self-consistent  Hartree and Hartree-Fock
approximations have been widely used to
study the spectra of quantum dots.\cite{Alhassid} One expects
the HF approximation to be better suited for systems having large numbers of electrons
and states. \cite{Kunz} Several authors used the HF approximation
to calculate the occupation of levels on dots coupled to
reservoirs. Sindel {\it et al.} \cite{SINDEL} considered a two-level spinless
quantum dot, and found agreement between the results of the
Hartree approximation and those of the  numerical renormalization-group for
levels whose width is small compared to their separation. When that
width is comparable to (or larger than)  the level spacing the
charging of a level has been found to oscillate with the gate voltage.
Goldstein and Berkovits \cite{GOLDSTEIN} used the HF approximation
to study  a multi-level quantum dot. They also  obtained a
non-monotonic behavior of the level charging when one level was much broader than
the others; but when the width of the levels was small compared to
their spacing the level charging was found  to be monotonic as a function of an applied gate voltage. Below we consider
many   resonant levels, whose width is small compared to their
spacing. We expect the HF approximation to be valid in that regime.
Indeed, our results for the level occupations are in agreement
with the results of Refs. ~\onlinecite{SINDEL} and
~\onlinecite{GOLDSTEIN}. In the opposite regime, where the width of the levels is much larger than the level spacing, the so-called open dot regime, solving a kinetic equation for the occupancy on the dot, Catelani and Vavilov \cite{CATELANI} have found corrections to the shot noise due to electronic interaction, which do not manifest themselves in the conductance, that depend non-trivially on bias, temperature, and interaction
strength.

The linear-response  ac conductance
of a multi-level quantum dot coupled to a {\it
single} lead was measured  by Gabellli \textit{et
al.}\cite{GABELLI}, and found to exhibit  a quantized behavior which
suggested the use of this device as an on-demand coherent
single-electron source. Expanding the ac conductance at low
frequencies, and using the Hartree \cite{BUTTIKER3} and
Hartree-Fock \cite{ZOHAR} approximations, the oscillations in the
conductance were calculated and found to be in agreement with the experimental
data. Here we extend these calculations to the case in which the multi-level dot is coupled to \textit{two} leads. We find that the dc conductance peaks broaden as the Fermi energy is elevated and that the energy difference between two consecutive peaks grows linearly as a function of the Coulomb interaction strength.

We begin in Sec.  \ref{DETCAL}   by defining the different  possible noise spectra, and
continue with a brief description of the scattering
formalism. We then  explain in detail the model
Hamiltonian and its  mapping onto an effective
Hamiltonian using a mean-field approximation. The results of the
self-consistent approximation and the  noise spectra are
presented in Sec. \ref{SEC NOISE}, and summarized and discussed in Sec. \ref{APPENDIX B}.
In view of the controversy mentioned above, we devote a great part of Sec. \ref{APPENDIX B} to analyze
the  dip appearing in the noise spectrum of a (non-interacting)
two-level dot, and study in particular its modifications under the
application of a bias voltage and finite (although low) temperatures.
In the  Appendix we explore the possible experimental detection of the
Fock terms.

\section{Details of the calculation }

\label{DETCAL}

\subsection{The noise spectra of a two-terminal system}

\label{NOISE}

Our calculation is carried out for a quantum dot connected to two electronic reservoirs,
denoted $L$ and $R$,  by single-channel
leads.
The reservoirs are characterized by
their chemical potentials, $\mu_{L}$ and $\mu_{R}$, respectively, such that the electronic
population in each of them is given by the Fermi distribution
\begin{align}
f^{}_{\alpha}(E )=\frac{1}{e^{(E -\mu_{\alpha})/k_{\rm B}T}+1}\ .\label{FD}
\end{align}
Below, we measure energies relative to the common Fermi energy,
\begin{align}
\epsilon^{}_{\rm{F}} =\frac{1}{2}(\mu_{L}^{}+\mu_{R}^{})\ .\label{EF}
\end{align}
The bias voltage $V$ across the quantum dot is
\begin{align}\label{Voltage}
V=(\mu^{}_L - \mu^{}_R)/e\ .
\end{align}

One may consider  two types of current-current correlations.
The correlation of the net current,
\begin{align}
 \label{Net Current}
\hat{I} = (\hat{I}^{}_L - \hat{I}^{}_R)/2\ ,
\end{align}
flowing through the system, $C^{(-)}(\omega )$, is [see
Eq. (\ref{Unsymmetrized noise})]
\begin{align}
\label{Neg Noise}
C^{(-)}(\omega)&\equiv \int _{-\infty} ^{\infty} dt e^{i\omega t}\langle
\delta \hat{I}_{}(t)\delta \hat{I}_{}(0) \rangle \nonumber\\
&=\frac{1}{2}\Bigl ( C^{({\rm auto})}(\omega) - C^{(\times)}(\omega) \Bigr )\ .
\end{align}
Here we have introduced the definitions
\begin{align}
C^{({\rm auto})}_{}(\omega) &= \frac{1}{2} \Bigl ( C^{}_{LL}(\omega)+C^{}_{RR}(\omega) \Bigr )\ ,\nonumber\\
C^{(\times)}_{}(\omega)& = \frac{1}{2} \Bigl ( C^{}_{LR}(\omega)+C^{}_{RL}(\omega)\Bigr )\ ,\label{CCR}
\end{align}
for the auto correlation and the cross correlation, respectively.
[Note that both $C^{({\rm auto})}(\omega)$ and $C^{(\times)}(\omega)$ are real.]
The other correlation function is associated with the rate by which charge is accumulated on the dot,
and is given by
\begin{align}
 C^{(+)}_{}(\omega)&\equiv \int _{-\infty} ^{\infty} dt e^{i\omega t}\langle
\Delta \hat{I}^{}_{}(t)\Delta \hat{I}^{}_{}(0) \rangle\nonumber\\
&=\frac{1}{2}\Bigl ( C^{({\rm auto})}(\omega) + C^{(\times)}(\omega) \Bigr )\ ,
\label{Pos Noise}
\end{align}
where
\begin{align} \label{Pos Current}
  \Delta \hat{I} = (\hat{I}^{}_L + \hat{I}^{}_R)/2\ .
\end{align}
As opposed to the net current operator Eq. (\ref{Net Current}), for which
$\langle  \hat{I} \rangle\neq 0$ in the presence of a finite bias voltage,
$\langle \Delta \hat{I} \rangle$=0. \cite{BLUNTER}
It was shown in Ref. ~\onlinecite{ME} that at zero frequency the noise
associated with $\Delta\hat{I}$ is always zero, i.e.,  $C^{(+)}(0)=0$.

\subsection{The noise spectra in the scattering formalism}

\label{SCAT}

As discussed in Sec. \ref{INTRO}, we treat the electronic
interactions  by the self-consistent Hartree-Fock approximation.
In this approximation the Hamiltonian becomes effectively a single-particle one, and therefore one may
use the (single-particle) scattering formalism \cite{BUTTIKER1,BUTTIKER2} to express the current
correlation functions, Eqs. {\eqref{Neg Noise}-\eqref{Pos Noise}, in terms of the elements of the scattering
matrix.

In the scattering formalism  the system is prresented by its scattering matrix;
when the system is coupled to two reservoirs by single-channel leads the latter is a
2$\times $2 matrix, denoted $S_{\alpha\alpha '}(E)$, which is  a
function of the energy $E$ of the scattered electron. One finds  \cite{BUTTIKER2}  (we employ units in which $\hbar=1$ hereafter)
 \begin{align}
  \label{AA'}
C^{}_{\alpha \alpha '}(\omega) &= \frac{e^2}{2 \pi} \int _{-\infty}
^{\infty} dE  \sum _{\gamma \gamma '}
F^{\alpha\alpha'}_{\gamma\gamma'}(E,\omega)\nonumber\\
&\times f^{}_{\gamma} (E )\Bigl  (1 - f^{}_{\gamma '} (E+\omega)\Bigr )\ ,
\end{align}
where $f$ is the Fermi function, Eq. (\ref{FD}), and
\begin{align}
F^{\alpha\alpha'}_{\gamma\gamma'}&(E,\omega)\equiv A ^{}_{\gamma
\gamma '} (\alpha,E ,E+\omega) A^{}_{\gamma'
\gamma} (\alpha', E+\omega, E)\ ,
\end{align}
with
\begin{align}
A^{}_{\gamma\gamma '}(\alpha , E, E') =\delta^{}_{\gamma\gamma
'}\delta^{}_{\alpha \gamma}-S^{\ast}_{\alpha\gamma}(E)S^{}_{\alpha
\gamma '}(E')\ .\label{AA}
\end{align}
The various correlation functions
are each given as  a sum of four processes: the intra-lead contributions, which include the combinations
$f_{L}(E )[1-f_{L}(E+\omega)]$ and
$f_{R}(E )[1-f_{R}(E+\omega)]$,
and the inter-lead ones, which contain
$f_{L}(E )[1-f_{R}(E+\omega)]$
and $f_{L}(E )[1-f_{R}(E+\omega)]$.
The relative contribution of each process to
$C^{(\pm)}(\omega )$ is determined
by the respective product of the Fermi functions.
In particular, at $T=0$, this product vanishes everywhere except
in a finite segment of the energy axis, determined by the bias voltage
[see Eq. (\ref{Voltage})]  and the frequency $\omega$. Finite temperatures
broaden and smear the limits of this segment, while varying
the bias voltage may shift it along the energy axis
or change its length.

\subsection{The self-consistent Hartree-Fock approximation}

\label{MODEL}

Here  the model Hamiltonian we use is specified, and
the self-consistent Hartree-Fock approximation by which  it  is reduced    to an effectively
single-particle one is described. That effective Hamiltonian is employed to derive   the scattering matrix.

Our model system consists of
 a multi-level quantum dot coupled to two (spinless) electronic reservoirs via two single-channel leads.
 The Hamiltonian of such a system is
\begin{align}
 \label{H}
 \mathcal{H} = \mathcal{H}^{}_{leads} + \mathcal{H}^{}_{tun} + \mathcal{H}^{}_{dot}\ ,
\end{align}
where the leads are described by
\begin{align}
\label{hleads}
 \mathcal{H}^{}_{leads}= \sum_{\alpha =L,R}\sum_{k} \epsilon^{}_{k\alpha} c_{k\alpha}^{\dag} c^{}_{k\alpha} \ ,
\end{align}
with
$\epsilon_{k\alpha}$ being the energy of an electron in  lead $\alpha$, and
$c_{k\alpha}^{\dag}$ ($c^{}_{k\alpha}$) is a creation (annihilation)
operator of an electron with momentum $k$ in lead $\alpha$.
The tunneling part of the Hamiltonian is
\begin{align}
{\cal H}^{}_{tun}=\sum_{\alpha=L,R}\sum_{k,n}
\Bigl (V^{}_{k\alpha n}d^{\dagger}_{n}c^{}_{k\alpha}+hc\Bigr )\ ,\label{Htun}
\end{align}
where $V_{k\alpha n}$ is the tunneling probability amplitude of an
electron to hop from lead $\alpha$ to the $n$th
level  on the quantum dot [$d_{n}^{\dag}$ ($d^{}_{n}$) is a creation
(annihilation) operator of an electron on that  level]. The Hamiltonian of the dot is
\begin{equation}\label{Hdot}
 \mathcal{H}_{dot} ^{}= \sum_{n} \epsilon^{}_n d^{\dag} _n d^{}_n
+ \frac{U}{2} ( \sum_{n}  d^{\dag} _n d^{}_n-{\cal N}^{}_g)^2\ ,
\end{equation}
where  $\epsilon_n$ is the single-particle energy on the $n$th level of the dot.
The electronic interactions (taking place solely on the dot) are described by the last term of Eq. (\ref{Hdot}),
adopting the ``orthodox" form,  \cite{MATVEEV}
in which
$U\equiv e^2/C$
represents the capacitive charging energy ($C$ is associated with the capacitance of the dot) and ${\cal N}_{g}$
is proportional to the gate voltage.
The role played by this parameter is discussed below.

As is mentioned above,
the interactions  are treated  within the framework of the self-consistent Hartree-Fock approximation \cite{BRUUS}.
This is accomplished as follows. The product of four
operators appearing in Eq. (\ref{Hdot}), assuming $n \neq m$,  is replaced by
\begin{align}
 \label{MF}
 d^{\dag}_n d^{}_n d^{\dag}_m d^{}_m \rightarrow \ & d^{\dag}_n d^{}_n  \langle d^{\dag}_m d^{}_m \rangle
  + \langle d^{\dag}_n d^{}_n \rangle d^{\dag}_m d^{}_m \nonumber\\
  &-  \langle d^{\dag}_n d^{}_m  \rangle d^{\dag}_m d^{}_n
        -  d^{\dag}_n d^{}_m \langle d^{\dag}_m d^{}_n \rangle\nonumber
\\ &-  \langle d^{\dag}_n d^{}_n \rangle\langle d^{\dag}_m d^{}_m\rangle + |\langle d^{\dag}_n d^{}_m \rangle|^2\ .
\end{align}
The decomposition given by the first two terms on the right-hand side
of Eq.  (\ref{MF}) represents the Hartree approximation \cite{BRUUS}
(these are also referred to as ``diagonal terms") while the next two terms represent the
Fock one (and are also referred to as ``non-diagonal terms").
The Fock terms were discarded in some of
the  previous works \cite{BUTTIKER3},   though in Ref. ~\onlinecite{ZOHAR} they were shown
to be of paramount importance. We show below that this is not the case in our calculation.

In the Hartree-Fock approximation, the dot Hamiltonian
(now denoted  $\widetilde{\cal H}_{dot}$)  becomes effectively a single-particle one,
\begin{align}
\label{hdoteff}
\widetilde{\mathcal{H}}_{dot}^{}= \sum_{n}\widetilde{\epsilon}^{}_n d^{\dag}_n d^{}_n -
\sum _{ \stackrel{n,m}{n \neq m}}\widetilde{J}_{nm}^{}d^{\dag}_n d^{}_m \ ,
\end{align}
where constant terms have been omitted. In this form of the Hamiltonian, the single-particle energy, $\widetilde{\epsilon}_{n}$, is
\begin{align}
\label{effective parameters e}
\widetilde{\epsilon}_n = \epsilon_n + U\Bigl ( \sum_{m\neq n} \langle d^{\dag}_m d^{}_m \rangle  -{\cal N}^{}_g\Bigr )\ ,
\end{align}
while the effective hopping amplitudes, $\widetilde{J}_{nm}$,  are
\begin{align}
\label{effective parameters J}
\widetilde{J}_{nm} ^{}=  U\langle d^{\dag}_m d^{}_n \rangle\   .
\end{align}
These new parameters of the Hamiltonian are calculated self-consistently.
As our system possesses time-reversal symmetry, the effective hopping amplitudes
$\widetilde{J}_{nm}$ can be assumed to be real (see below).

For the sake of clarity, we confine ourselves  to very low temperatures and to the linear-response regime, and consequently
carry out the self-consistent calculation of  $\widetilde{\epsilon}_{n}$
and $\widetilde{J}_{nm}$ at  zero bias, using the relation
\begin{align}
\langle d^{\dagger}_{n}d^{}_{m}\rangle =\frac{i}{2\pi}\int_{-\infty}^{\infty}dE f(E)\Bigl [{\cal G}^{r}_{mn}(E)-\Bigl ({\cal G}^{r}_{nm}(E )\Bigr )^{\ast}_{}\Bigr ]\ .\label{Qnm}
\end{align}
The retarded Green function matrix of the dot,
\begin{align}
{\cal G}^{r}_{nm}(E)=\Bigl [E-\widetilde{\cal H}^{}_{dot}-\Sigma^{r}_{}(E)\Bigr ]^{-1}_{nm}\ , \label{GREEN}
\end{align}
is expressed in terms of the effective Hamiltonian (\ref{hdoteff}), and the self-energy matrix
$\Sigma^{r}$  due to the coupling with the leads is given by
\begin{align}
\Sigma^{r}_{nm}(E)=\sum_{\alpha}\sum_{k}\frac{V^{}_{k\alpha n}V^{\ast}_{k\alpha m}}{E -\epsilon^{}_{k\alpha}+i\eta}\ ,
\label{SIGMAR}
\end{align}
with $\eta\rightarrow 0^{+}$. In Eq. (\ref{Qnm}) $f(E)$ denotes the {\em thermal-equilibrium } Fermi function of the leads obtained from Eq. (\ref{FD}) by using  $\mu_{\alpha}=\epsilon_{\rm{F}}$.
As usual, $-{\rm Im}\Sigma^{r}_{nn}(E)$ describes the broadening of the energy levels on the dot due to the coupling with the leads.

\subsection{The scattering matrix and the self-consistent calculation}
\label{SEC SELF-CONSISTENT CALC}

In order to obtain an explicit expression for the scattering matrix, it is convenient to present the tunneling Hamiltonian [see Eq. (\ref{Htun})] in the site representation,  modeling each of  the leads by  a one-dimensional tight-binding chain of zero on-site energies and equal nearest-neighbor hopping amplitudes $-j$ (for simplicity, we assume the two leads to be identical). The coupling of the left (right) lead to the $n$th level on the dot is denoted $-V_{Ln}$  ($-V_{Rn}$). One then has $V_{kLn}=-\sqrt{2/N}V_{Ln}\sin k$ and  $V_{kRn}=-\sqrt{2/N}V_{Rn}\sin k$, where each of the two tight-binding chains consists of $N$ sites whose  lattice spacing is taken as unity. It follows that the scattering matrix at energy $E=-2j\cos k$ is
\begin{align}
S(E)=-1+\frac{iv(E)}{j^{2}}\left [\begin{array}{cc}V^{}_{L}{\cal G}^{r}_{}(E)V^{\dag}_{L}&V^{}_{L}{\cal G}^{r}_{}(E)V^{\dag}_{R}\\
V^{}_{R}{\cal G}^{r}_{}(E)V^{\dag}_{L}&
V^{}_{R}{\cal G}_{}^{r}(E)V^{\dag}_{R}\end{array}\right ]\ ,\label{SCATM}
\end{align}
where $V_{L}$ ($V_{R}$) is the row vector of coupling amplitudes of the dot levels to the left (right) lead consists of vector element $V_{L/R n}$, and $v(E)=2j \sin k$
is the group velocity of the scattered electron. Time-reversal symmetry dictates a {\em symmetric} scattering matrix. This in turn implies that the Green function is symmetric in the indices $n$ and $m$, and consequently $\langle d^{\dagger}_{m}d^{}_{n}\rangle $ [see Eq. (\ref{Qnm})] must be real.

In the following, we ignore the real part of the self energy of the dot Green function [see Eqs. (\ref{GREEN}) and
(\ref{SIGMAR})].  This is a plausible approximation provided that the real part does not vary significantly with the   energy. Then, the  (imaginary part of the) self energy consists of the coupling of the dot levels to the left and to the right lead, and reads
\begin{align}
&\Sigma^{r}_{}(E)=\frac{-i}{2}\Bigl (\Gamma^{}_{L}(E)+\Gamma^{}_{R}(E)\Bigr )\ ,\label{SIGR}
\end{align}
where the (symmetric) matrices $\Gamma_{L(R)}$ are given by
\begin{align}
\Gamma^{}_{L(R)}(E)=\frac{v(E)}{j^{2}}V^{\dag}_{L(R)}\otimes V^{}_{L(R)}\ .\label{GAMALR}
\end{align}
Employing these expressions in Eq. (\ref{Qnm}) leads to
\begin{align}
&\langle d^{\dagger}_{n}d^{}_{m}\rangle =\int_{-\infty}^{\infty} \frac{dE}{\pi}f(E)
\Bigl [{\cal G}^{r}_{}(E)i\Sigma^{r}_{}(E){\cal G}^{a}_{}(E)\Bigr ]^{}_{nm}\ ,\label{SC}
\end{align}
with ${\cal G}^{a}_{nm}(E)={\cal G}^{r*}_{nm}(E)$.
One may estimate the magnitudes of the averages $\langle d^{\dagger}_{n}d^{}_{m}\rangle$  by considering Eq. (\ref{SC}) in the limit of weak couplings to the leads, $\bar{\Gamma}/j\ll 1$, where  $\bar{\Gamma}\simeq v V^{2}/j^{2}$ denotes the typical value of the broadening of the dot levels due to that coupling. It is then expedient to use  the Lehmann representation for the Green function,
\begin{align}
{\cal G}^{\stackrel{r}{a}}_{}(E)&\simeq \Bigl (E -\widetilde{H}^{}_{dot}\pm i\eta \Bigr )^{-1}\nonumber\\
&=\sum_{\ell}\frac{|\psi^{}_{\ell}\rangle\langle\psi_{\ell}^{}|}{E \pm i\eta -\widetilde{E}^{}_{\ell}}\ ,\label{DECO}
\end{align}
where the
$\widetilde{E}_{\ell}$'s are the eigenvalues of $\widetilde{H}_{d}$ and the $|\psi_{\ell}\rangle $'s  are the corresponding eigenvectors. Notably,
the  $\widetilde{E}_{\ell}$'s differ from the original onsite energies on the dot, $\epsilon_{n}$, because of the Hartree corrections [see Eq.
(\ref{effective parameters e})] and due to the Hartree-Fock ones [see Eq.
(\ref{effective parameters J})], which render the dot Hamiltonian non diagonal.
However, when the coupling to the leads is small,
the averages
$\langle d^{\dagger}_{n}d^{}_{m}\rangle$ for $n\neq m$ are small as well, bringing $\widetilde{E}_{\ell}$  close to one of the original onsite energies and the eigenvector $|\psi_{\ell}\rangle$ to be close to the corresponding original basis vector for the noninteracting system. Inserting Eq. (\ref{DECO}) into Eq. (\ref{SC}) then yields
\begin{align}
\label{weak coupling}
&\langle d^{\dagger}_{n}d^{}_{m}\rangle \simeq
\frac{1}{\pi }\int _{-\infty}^{\infty}dE f(E)\nonumber\\
&\times\sum_{\ell_{1}\ell_{2}}\Biggl [\frac{|\psi^{}_{\ell_{1}}\rangle\langle\psi^{}_{\ell_{1}}|\Bigl (\Gamma^{}_{L}(E) +\Gamma^{}_{R}(E) \Bigr )|\psi^{}_{\ell_{2}}\rangle\langle\psi^{}_{\ell_{2}}|}{(E -\widetilde{E}^{}_{\ell_{1}}+i\eta )(
E-\widetilde{E}^{}_{\ell_{2}}-i\eta)}\Biggr ]^{}_{nm}\ .
\end{align}
Examination of this expression (in the zero-temperature limit) reveals that in most cases the result of this integral is rather small, of the order of $\bar{\Gamma}/|\epsilon_{n}-\epsilon_{m}|$. A significant result is obtained only when $n=m$ (and consequently $\ell_{1}\simeq\ell_{2}$).

These observations are borne out by numerically solving the self-consistent equations (\ref{SC}).
The computations have been carried out for a dot including 61
equally-spaced levels, $\epsilon_{n}=n\Delta$ ($n=-30,...,30$),
which are all  coupled identically to each of the  two leads. \footnote{The value of $61$ levels was used in Refs. ~\onlinecite{BUTTIKER3} and ~\onlinecite{ZOHAR}. As they explain, the HF approximation is better for a large number of levels.}
That coupling  [see Eq. (\ref{Htun})] has been modeled by a saddle-like
point contact-potential,  \cite{BEENAKKER2}
\begin{align}
\Gamma^{}_{\alpha}(\epsilon^{}_{\rm{F}})\equiv\frac{v(\epsilon^{}_{\rm{F}})}{j^{2}}V^{2}_{\alpha}\rightarrow\frac{\Delta}{\pi \mathcal{T}^{}_{\alpha}(\epsilon^{}_{\rm{F}})}\Bigl (1-
\sqrt{1-\mathcal{T}^{}_{\alpha}(\epsilon^{}_{\rm{F}} )}\Bigr )^2\ ,\label{BEEN}
\end{align}
where $\Delta$ is the level spacing, and $\alpha =L$ or $R$.
The saddle-like point contact is characterized by its transmission probability,
 \cite{BUTTIKER4}
 \begin{align}
 \label{T}
{\cal T}^{}_{\alpha}(\epsilon^{}_{\rm{F}})=\frac{1}{1+e^{-(\epsilon^{}_{\rm{F}}-\epsilon^{}_{\alpha})/\omega^{}_{\alpha}}}\ ,
\end{align}
where $\omega_{\alpha}$ and $\epsilon_{\alpha}$ are the parameters characterizing each of the point contacts.

\begin{figure}
 \includegraphics[width=9cm, height=6cm]{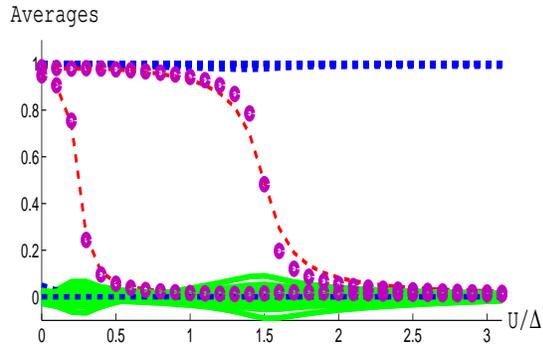} \\ \includegraphics[width=9.5cm, height=6cm]{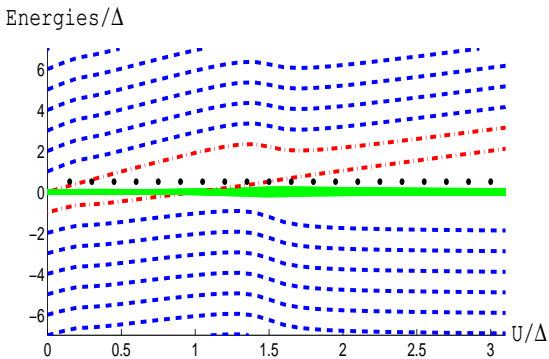}
 \caption{(color online) The top panel depicts the  averages, computed self-consistently from Eq. \eqref{Qnm},   as a function of the interaction strength $U$. The horizontal  curves  are the diagonal averages,
$\langle d_n ^{\dag}d_n \rangle$; the ``cloud" around the lowest ones represents   all  the off-diagonal averages $\langle d_n ^{\dag}d_m \rangle$ with $n\neq m$. The red (thin-dashed) step-like curves represent the diagonal average corresponding to the energy levels crossing the Fermi energy $\epsilon_{\rm{F}}$,  computed within the Hartree-Fock approximation;  the purple (thick-dotted) curves show the same average computed within the Hartree approximation alone. Lower panel:  the effective energies $\widetilde{\epsilon}_{n}$, Eq.
(\ref{effective parameters e}), are presented by the blue (dashed) and red (dashed-dotted) curves. The green (continuous) curves are $\widetilde{J}_{nm}$, Eq. (\ref{effective parameters J}), and the dotted (black) horizontal curve marks the Fermi energy. Parameters: $\epsilon_{\rm{F}}=0.5$ ; $\epsilon_L=\epsilon_R=0.598$ ; $\omega_L=\omega_R=0.8$ ; ${\cal N}^{}_g=30$. The level widths [see Eqs. \eqref{BEEN} and \eqref{T}] are then equal,  $\Gamma_L=\Gamma_R=0.05$. (All energies are in units of the level spacing $\Delta$.)}
\label{AVER}
\end{figure}

Representative  results of the self-consistent computations are portrayed in Figs.  \ref{AVER} and \ref{AVER1}, as a function of the interaction $U$.  To produce those curves, we have chosen parameters such that the energy levels on the dot remain well-separated also after being modified by the Hartree terms and the Fermi energy is chosen such that for $U=0$ the energy levels are either occupied or unoccupied. In the lower panel of Fig \ref{AVER} we see that lines representing the effective energy levels, Eq. \eqref{effective parameters e}, start linearly, since $\langle d^\dag _n d_n  \rangle$ are either zero or one, as shown in the top panel of Fig. \ref{AVER}. As soon as a level crosses the Fermi level (the $n=0$ level), its occupation change from $1$ to $0$ (see the top panel) and as a consequence the linear slope of the effective energies changes. After the second level (the $n=1$ level) crosses the Fermi level the levels below the Fermi level return the their non-interacting values and the levels above the Fermi level grow linearly with $U$. The parameter $\mathcal{N}_g$ was chosen to be $30$ in Fig \ref{AVER}. We find that in this parameter range, the Fock averages are negligible almost always. It is only when the $n$th  modified level, $\widetilde{\epsilon}_{n}$, crosses the Fermi energy  (upon varying $U$) that the non-diagonal averages  $\langle d^{\dag}_{n+i} d^{}_{n+i'}\rangle$, of small integer values  $i$ and $i'$,  acquire  significant values. A further investigation of the Fock terms is given in the Appendix for a two-level dot.

The only parameter distinguishing Fig. \ref{AVER1} from Fig. \ref{AVER} is  ${\cal N}_g$, taken to be $33$ in the former. The parameter ${\cal N}^{}_g$ determines the number of levels below the Fermi energy at large interaction. In Fig. \ref{AVER1} we see a similar behavior to the one in Fig. \ref{AVER} only that now the effective energy levels have a negative slope for small values of $U$ (as shown in the lower panel) and that the total number of occupied levels in the dot increases as a function of $U$ (as shown in the top panel).  Denoting the total number of energy levels on the dot by $2n_{d}+1$, we find that
for ${\cal N}^{}_g-n_d>\epsilon_{\rm{F}}/\Delta$ [${\cal N}^{}_g-n_d<\epsilon_{\rm{F}}/\Delta$], the modified energy levels  shift as a function of $U$ such that at large $U$'s there will be mod$({\cal N}^{}_g)$ [mod$({\cal N}^{}_g)+1$] levels below the Fermi energy. This is indeed the case in Figs. \ref{AVER} and \ref{AVER1}.

\begin{figure}
\includegraphics[width=9cm, height=6cm]{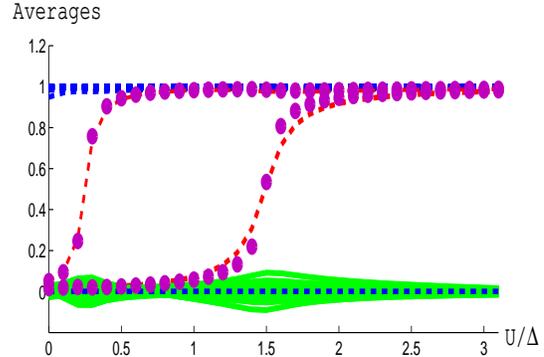} \\ \includegraphics[width=9.5cm, height=6cm]{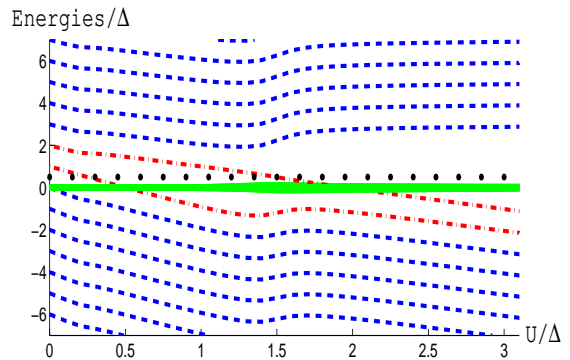}
\caption{(color online) Same as Fig. \ref{AVER}, with ${\cal N}^{}_g=33$.}
\label{AVER1}
\end{figure}

\section{The noise spectrum}
\label{SEC NOISE}

As the self-consistent determination of the effective dot Hamiltonian,
Eq. (\ref{hdoteff}),  is strictly-speaking valid only at zero bias and zero temperature, we present
in this section the noise spectra calculated in this regime. We comment on
the effects caused by a finite bias voltage and finite (but low) temperatures  in  Sec.  \ref{APPENDIX B}.
Our computations are confined to the case in which all levels on the dot are coupled identically to each
of the two leads.  In that case, the scattering matrix takes the simple form [see Eqs. (\ref{SCATM}) and (\ref{GAMALR})]
\begin{align}
\label{SCATg}
&S(E)=-1\nonumber\\
&+ig^{r}_{}(E)\left [ \begin{array}{cc}
 \Gamma_{L}^{} (E)& \Gamma_{L}^{1/2}(E) \Gamma_{R}^{1/2}(E) \\
\Gamma_{L}^{1/2}(E) \Gamma_{R}^{1/2}(E) & \Gamma_{R}^{}(E) \\
\end{array}
\right ]\ ,
\end{align}
where
\begin{align}
g^{r}_{}(E)=\sum_{n,m} {\cal G} ^r _{nm}(E)\label{LITG}
\end{align}
depends on the sum of the two widths [see Eq.
(\ref{SIGR})]
\begin{align}
\Gamma (E)=\Gamma^{}_{L}(E)+\Gamma^{}_{R}(E)\ .
\end{align}
Furthermore, since the computations below are confined to the linear-response regime,
$\Gamma_{L}(E)$ and $\Gamma_{R}(E)$ are replaced by their values at the Fermi
energy, as given in Eqs. (\ref{BEEN}) and (\ref{T}). It therefore follows that
\begin{align}
g^{r}_{}(E)-g^{a}_{}(E)=-i\Gamma g^{r}_{}(E)g^{a}_{}(E)\ .\label{DIFG}
\end{align}

Using Eqs. (\ref{SCATg}) and (\ref{DIFG})  in conjunction with Eqs. (\ref{CCR}) yields
\begin{align}
&C^{({\rm auto})}_{}(\omega )=\frac{e^{2}}{4\pi}\int_{-\infty}^{\infty}dEf(E)[1-f(E+\omega)]\nonumber\\
&\times \Bigl (\Gamma^{2}_{}[g^{r}_{}(E)|^{2}+|g^{r}_{}(E+\omega )|^{2}]\nonumber\\
&-[\Gamma^{2}_{L}+\Gamma^{2}_{R}][g^{r}_{}(E)g^{a}_{}(E+\omega )+{\rm cc}]\Bigr )\ ,\label{CAUTO}
\end{align}
and
\begin{align}
C^{(\times)}_{}(\omega )=&-\frac{e^{2}}{2\pi}\int_{-\infty}^{\infty}dEf(E)[1-f(E+\omega)]\nonumber\\
&\times\Gamma^{}_{L}\Gamma^{}_{R}\Bigl (g^{r}_{}(E)g^{a}_{}(E+\omega )+{\rm cc}\Bigr )\ ,\label{CTIM}
\end{align}
from which the two correlations introduced
in Sec. \ref{SCAT}, $C^{(-)}$ [Eq. (\ref{Neg Noise}) for the net current] and $C^{(+)}$ [Eq. (\ref{Pos Noise}), for the charge fluctuations]
can be derived.
Interestingly enough, the latter correlation  depends only on the sum of the two widths, $\Gamma$, and therefore is insensitive
to the (possible) asymmetry of the couplings of the dot to the leads.

We begin our analysis with a discussion of the zero-temperature dc conductance, as
given by the Landauer formula
\begin{align}
\label{DC}
  G=\frac{e^2}{2\pi}{\cal T}(\epsilon_{\rm{F}}) \ ,
\end{align}
with the transmission coefficient ${\cal T}(\epsilon_{\rm{F}})=\Gamma_L\Gamma_R|g^r(\epsilon_{\rm{F}})|^2$.
 In the absence of the interaction, the conductance
as a function of the Fermi energy exhibits peaks whenever the Fermi energy coincides
with one of the levels of the dot, see the solid line in  Fig. \ref{FIGCOND}. The peaks of the conductance broaden as the
Fermi energy increases; this results from the energy dependence of the coupling between the
leads and the dot, modeled by point contacts [see Eqs. \eqref{BEEN} and \eqref{T}]. When the interactions on the dot are taken into account (in the self-consistent HF
approximation) the levels on the dots are shifted and so are the conductance
peaks (see the dashed curve in Fig. \ref{FIGCOND}). Moreover,  when the transmission
probability of each of the point contacts is small,  the
dot is in the so-called Coulomb Blockade regime. This blockade can be lifted by
tuning the  Fermi energy or the gate voltage (equivalent to the parameter ${\cal N}_{g}$ in our model). In  our case,
the lifting occurs when the Fermi energy of the electrons in the leads coincides with
one of the onsite levels on the dot; this happens at energy intervals of size
$\Delta+U$, and is associated with the addition of one electron to the dot.

\begin{figure}[ht]
\includegraphics[width=9cm]{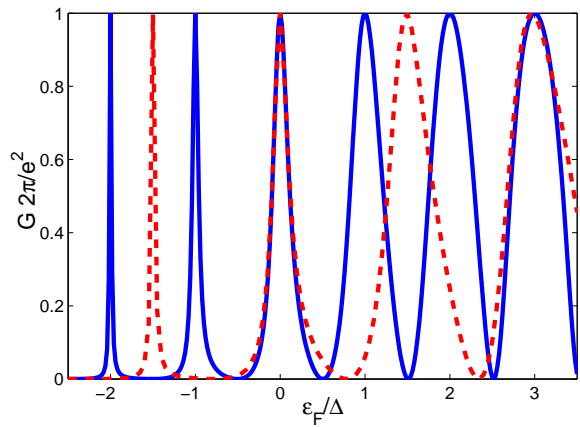}
\caption{(color online) The dc conductance as a function of the Fermi energy. The continuous (blue) curve is the conductance of a non interacting dot, Eq. \eqref{DC}; the dashed (red) curve corresponds to  $U=0.5$. Parameters used in Eq. \eqref{T}: $\omega_L=\omega_R=0.8$ ; $\epsilon_L=\epsilon_R=0$. (All energies are in units of the level spacing $\Delta$.)}
\label{FIGCOND}
\end{figure}

\begin{figure}[ht]
\includegraphics[width=9.cm]{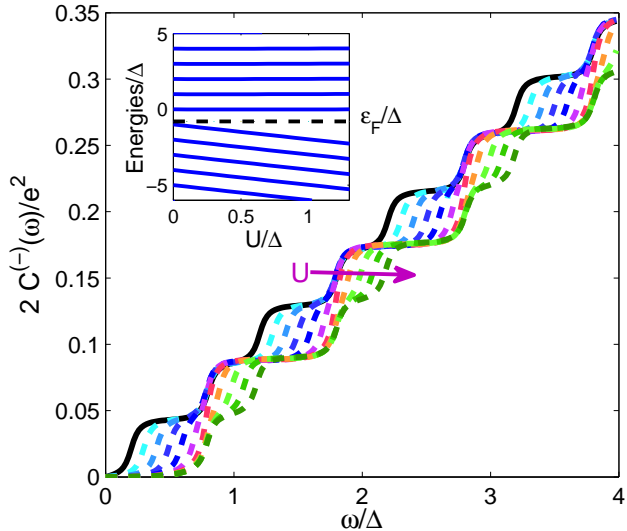}
\caption{(color online) The noise spectrum for a dot coupled symmetrically to the leads, as a function of the  frequency, for various values of the
interaction strength ($U=0,$ 0.1,$\dots$ , $1$, from left to right) computed within the Hartree approximation.
Inset: the effective single-particle
energy levels $\widetilde{\epsilon}_n/\Delta$ on the dot, as a function of the interaction strength for the same parameters. Parameters:
$\epsilon_F = -0.8$, ${\cal N}^{}_g=30$ and $\Gamma_L=\Gamma_R=0.025$. These widths are given by Eqs. (\ref{BEEN}) and  (\ref{T}) with  $\epsilon_L=\epsilon_R= 0$ and $\omega_L=\omega_R=0.8$.  (All energies are in units of the level spacing $\Delta$.)}
\label{FIGNOISEU}
\end{figure}

Turning now to the noise spectra, we portray below our results for the net current correlation, Eq. (\ref{Neg Noise}).
Figure \ref{FIGNOISEU} shows the noise spectra (as a function of the frequency $\omega$) for various values of the interaction strength,
for a dot coupled {\em symmetrically} to both leads.
The left-most curve in the main panel  depicts the correlation in the interaction-free case. The spectrum shows a series of steps, of height $\Gamma$ (in units of $e^{2}$), located roughly at frequencies corresponding to the renormalized onsite energies on the dot. As the interaction strength is increased the levels are shifted due to the Hartree correction. (Those effective onsite energies are depicted in the inset of Fig. \ref{FIGNOISEU}.)  As a result, the spectra are shifted as well. This step structure is similar to the ones obtained before, in the case of a non interacting  two-level \cite{STEPS AND DIPS} and  single-level \cite{ME} dots.
The curves in Fig. \ref{FIGNOISEU} are calculated within the Hartree approximation alone; including the Fock terms almost does not change them (see below).

\begin{figure}[h]
\includegraphics[width=11cm]{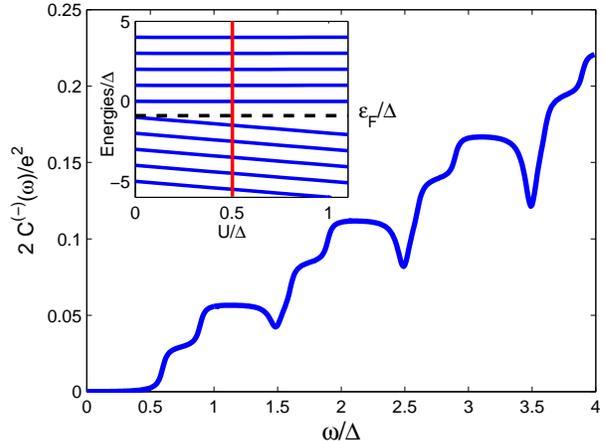}
\caption{(color online)
Same as Fig. \ref{FIGNOISEU} for a dot coupled asymmetrically to the leads. The vertical red line in the inset indicates the chosen interaction strength, $U=0.5$, and the dashed black line indicates the Fermi energy, $\epsilon_F = -0.9$. The dot lead coupling parameters are: $\Gamma_{L} =0.022$ and  $\Gamma_R=0.014$. The widths are found from Eqs. (\ref{BEEN}) and (\ref{T}) with $\epsilon_L=0$, $\epsilon_R=1$  and $\omega_L=\omega_R=0.8$.}
\label{FIGNSYMNOISE}
\end{figure}

When  the coupling of the dot to the leads is not symmetric, i.e., $\Gamma_{L}\neq \Gamma_{R}$, there also appear dips in-between the steps, as seen in Fig. \ref{FIGNSYMNOISE}. The frequencies at which the dips are located
correspond to the {\em spacing} between any two levels that happen to be on the two sides of the Fermi energy, as seen in the inset of Fig. \ref{FIGNSYMNOISE} where the effective energies are plotted as a function of the  interaction. We discuss further the origin of the dip structure in  Sec.  \ref{APPENDIX B}.

As mentioned above, we have chosen our parameters such that the onsite levels on the dot remain well-separated also when modified by the interaction; as a result of this choice, the Fock terms' contribution is almost negligible, and therefore the results shown in Figs. \ref{FIGNOISEU}
and \ref{FIGNSYMNOISE} were computed within the Hartree approximation alone. To further clarify this point, we compare  in Fig.
\ref{FIGNNOISEHF} the correlation $C^{(-)}(\omega) $ obtained in the interaction-free case with the one computed by the Hartree approximation, and within the Hartree-Fock  one.  As can be seen in that figure, the last two curves are almost indistinguishable.

\begin{figure}[ht]
\includegraphics[width=9cm,height=6.8cm]{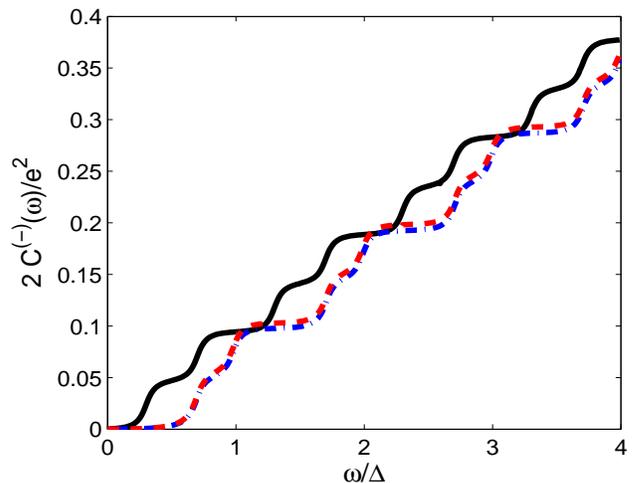}
\caption{(color online) Same as Fig. \ref{FIGNOISEU} for an interaction-free dot, shown by the solid (black) curve, and for $U=0.7$ computed within the Hartree approximation, shown by the dotted-dashed (blue) curve, and within the Hartree-Fock approximation, shown by the dashed (red) curve. The Fermi energy is $\epsilon_F = -0.7$.}
\label{FIGNNOISEHF}
\end{figure}

Notwithstanding the significance of studying the noise spectra as a function of the interaction strength, it seems more practical to explore it when the Fermi energy (or  the lead chemical potential at finite, but low, temperatures) is varied. This dependence is presented in Fig.
\ref{FIGNFERMI}, for an interaction-free dot. The dependence of the spectrum on the Fermi energy originates from the $\epsilon_{F}-$ variation of the point contact's transmission. As the latter is decreasing, the height of the step seen in the spectrum is decreasing. The location of the step in Fig. \ref{FIGNFERMI} corresponds to the difference between the Fermi energy and the closest energy level. We have chosen parameters for the Fermi energy in Fig. \ref{FIGNFERMI} such that the different steps will appear around the same frequency.

\begin{figure}[ht]
\includegraphics[width=9cm,width=8cm]{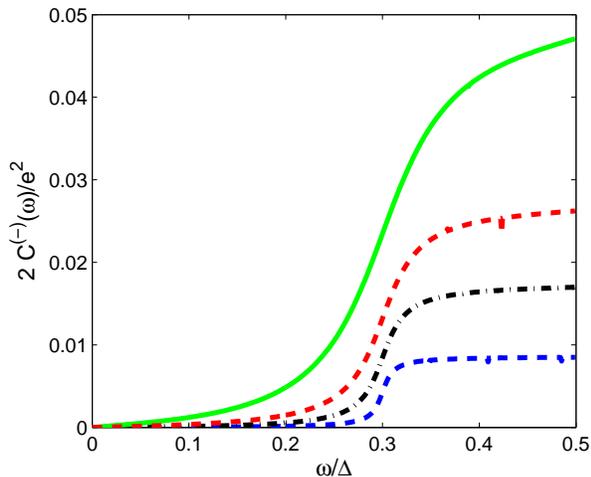}
\caption{(color online)  The noise spectrum of a non interacting dot as a function of the  frequency,  for various values of the Fermi energy,
$\epsilon_{F}=-0.7$, $-1.3$, $-1.7$, and $-2.3$ (from left to right).  Parameters:  $\Gamma_L=\Gamma_R=0.028$, $0.014$, $0.009$, $0.004$ [from left to right, obtained from Eqs. (\ref{BEEN}) and (\ref{T}) with $\omega_L=\omega_R=0.8$  and  $\epsilon_L=\epsilon_R= 0$].  (All energies are in units of the level spacing $\Delta$.)}
\label{FIGNFERMI}
\end{figure}

\section{Discussion}

\label{APPENDIX B}

We have presented  a calculation of the noise spectra of the currents passing through a multi-level  dot,
treating the Coulomb interactions (assumed to take place on the dot alone)
within the self-consistent Hartree-Fock approximation.
The computations
are carried out for the noise associated with the net-current fluctuations, as this quantity is
intimately related to the ac conductance.
Our main conclusion is that this correlation, as a function of the frequency, reflects the energy
spectrum of the dot. It shows steps at frequencies corresponding to the onsite energy levels of the dot,  and dips,
occurring at frequencies corresponding to the spacings between any two levels located
at the opposite sides of the Fermi energy.   The latter appear only when the dot is coupled
to the leads in an asymmetric form.
This pattern is similar to what has been obtained before \cite{STEPS AND DIPS,ME}
in the case of a few-level non interacting dots; this is not very surprising since we have
operated in the parameter regime where the Coulomb
interaction did not modify drastically the energy levels of the dot. For technical
reasons, the computations were restricted to zero temperature and zero bias voltage.

It is of course very desirable to examine the effects of finite bias voltages and finite
(though rather low) temperatures. In this section we carry out such a study,
confining ourselves to the simplest case of   a   two-level non interacting dot.
In the process, we also gain more insight into the origin of the step and dip pattern of the noise spectra.
In particular we focus on the dip structure since (as mentioned in Sec. \ref{INTRO})
it is debated in the literature.
We believe that the conclusions drawn from this simplistic model are valid, at least qualitatively,
also for the interacting multi-level dot, as long as the the interaction is not too strong
and the number of onsite levels is sufficiently large.

We begin with the   correlations $C^{(\pm)}$ at zero bias and zero temperature.
Then, Eqs. (\ref{CAUTO}) and (\ref{CTIM}) in conjunction with Eqs. (\ref{Neg Noise}) and (\ref{Pos Noise})  yield
\begin{widetext}
\begin{align}
C^{(\pm)}_{}(\omega )=\frac{e^{2}}{4\pi}\int_{\epsilon^{}_{F}-\omega}^{\epsilon^{}_{F}}dE\Bigl \{
(\Gamma^{}_{L}+\Gamma^{}_{R})^{2}\Bigl (|g^{r}_{}(E)|^{2}+|g^{r}_{}(E+\omega )|^{2}\Bigr )
-(\Gamma^{}_{L}\pm \Gamma^{}_{R})^{2}\Bigl (g^{r}_{}(E)g^{a}_{}(E+\omega )+{\rm cc}\Bigr )\Bigr \}\ .\label{CMINT}
\end{align}
\end{widetext}
In the case of a two-level, non interacting dot, the Green function Eq. (\ref{LITG}) takes the form
\begin{align}
\label{g}
g^{r}_{}(E)=\frac{(E-\epsilon^{}_1)+(E-\epsilon^{}_2)}{(E-\epsilon^{}_1
+i\frac{\Gamma}{2})(E-\epsilon^{}_2+i\frac{\Gamma}{2})+\frac{\Gamma^2}{4}} \ .
\end{align}
In particular, when the two levels are well-separated, $g^{r}$ consists roughly
of two Breit-Wigner resonances of (equal) widths $\Gamma/2$,
centered around $\epsilon_{1}$
and $\epsilon_{2}$.
The contribution of the first term in Eq. (\ref{CMINT}) to
the integration is hence quite clear:  As long as none of the resonances is within the integration
bounds the integral will almost vanish, while whenever the
frequency integration encompasses one of the Breit-Wigner
resonances, due to $g^r(E)$ [$g^r(E+\omega)$] in case the energy
level is below [above] the Fermi energy, the result will be a step in the
noise curve, whose height is roughly $\Gamma/2$ (in units of $e^2$).
The second term in the integrand of $C^{(-)}$ alone is non zero only when the dot is
coupled {\em asymmetrically} to the leads, namely, when $\Gamma_{L}\neq\Gamma_{R}$.
(This is not the case for the charge-related correlation $C^{(+)}$.) The integrand of this term consists of products
of resonances related to energy levels below and above the Fermi energy; the
contribution to the integration will be significant whenever the frequency, $\omega$, will roughly match the level spacing,
$\epsilon_{1}-\epsilon_{2}\simeq\omega$. This contribution is {\em negative},
and leads to the dip in the correlation as a function of the frequency of
depth roughly equal to $(\Gamma_L-\Gamma_R)^2/2\Gamma$ (in units of $e^2$).
Interestingly, the dip in $C^{(+)}$ suppresses the noise almost completely.

\begin{figure}[ht]
\includegraphics[width=8cm]{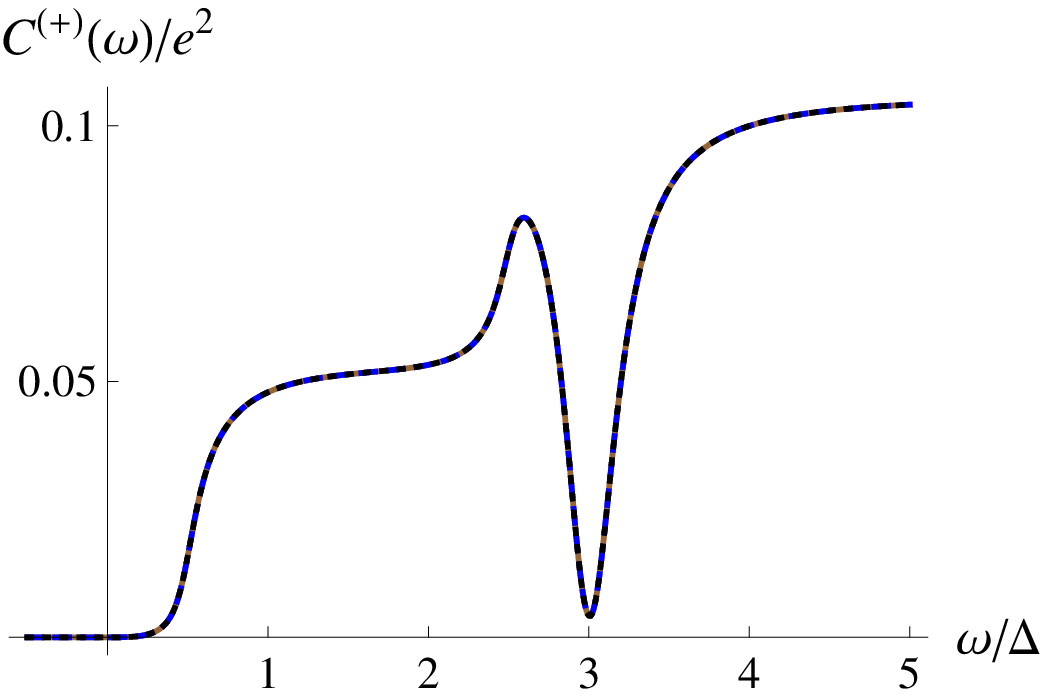}
\includegraphics[width=8cm]{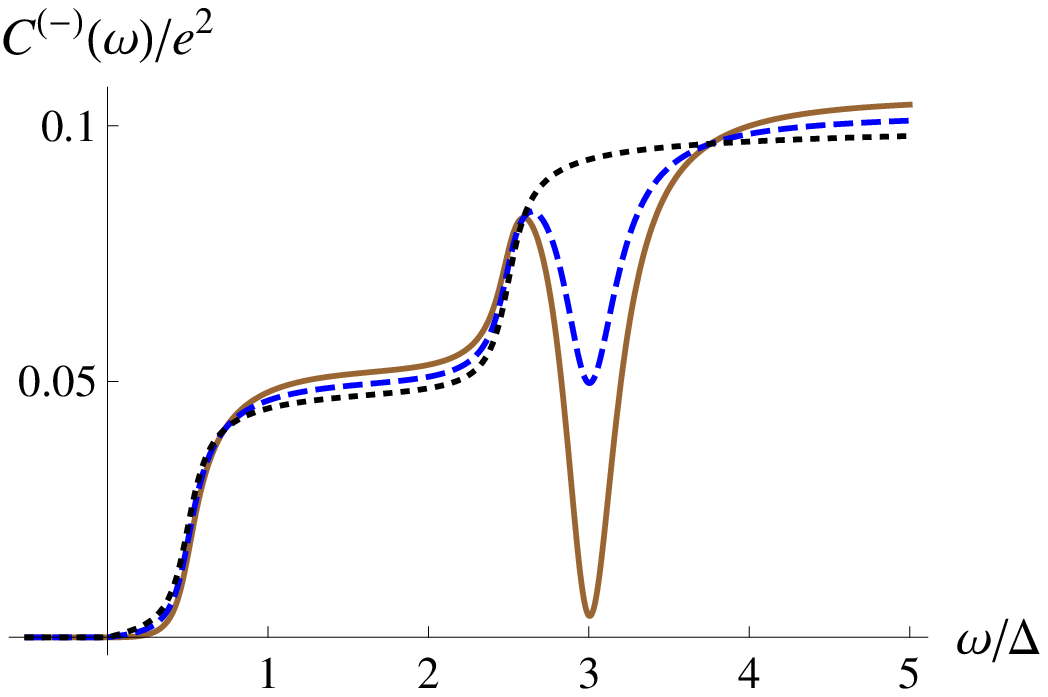}
\caption{(color online) The correlations $C^{(+)}$ (upper panel) and $C^{(-)}$ (lower panel) for the two-level case,
as a function
of the frequency at zero temperature and zero bias. The different curves in the
lower panel  correspond to $\Gamma_L=0.1$ and $\Gamma_R=0$ (brown
continuous curve), $\Gamma_L=0.0745$ and $\Gamma_R=0.0255$ (blue dashed curve), and $\Gamma_L=\Gamma_R=0.05$
(black dotted curve). Parameters: $\epsilon_{1}=1$, $\epsilon_{2}=-2$, and
$\epsilon_{F}=0.5$. All energies are in units of $\Delta$.}
\label{FIGCPM}
\end{figure}

Figure \ref{FIGCPM} depicts the behavior described above. In both correlations, $C^{(+)}$
and $C^{(-)}$, there appears a dip at the frequency corresponding to the difference
$\epsilon_{1}-\epsilon_{2}$; however, in the case of $C^{(-)}$,
that dip disappears once the dot is coupled symmetrically to the leads. In contrast,
the correlation $C^{(+)}$ is not sensitive to this asymmetry, as expected. The behavior
portrayed in Fig. \ref{FIGCPM} is in agreement with the findings of Refs.
~\onlinecite{STEPS AND DIPS} and ~\onlinecite{DONG},  but
contradicts those of  Ref. ~\onlinecite{MARCOS}, where a dip was obtained
even in the case where the dot is coupled symmetrically to the leads.
The two levels discussed in Ref. ~\onlinecite{MARCOS} are also coupled to
one another, making it a different system than ours.
The Green function $g^{r}$ [{\it c.f.} Eq. (\ref{LITG})] describing that geometry is
\begin{align}
\label{COUPLED g}
    g^{r}(E)=\frac{(E-\epsilon_1)+(E-\epsilon_2)-2J}{(E-\epsilon_1+i\frac{\Gamma}{2})(E-\epsilon_2+i\frac{\Gamma}{2})-(J+i\Gamma)^2} \ ,
\end{align}
where $-J$ is the coupling between the levels. Since the form of Eq. \eqref{CMINT} is
unaltered by the levels coupling [one has just  to employ Eq. (\ref{COUPLED g})] a
dip will not appear in the noise for $\Gamma_L=\Gamma_R$.

\begin{figure}
 \includegraphics[width=8cm]{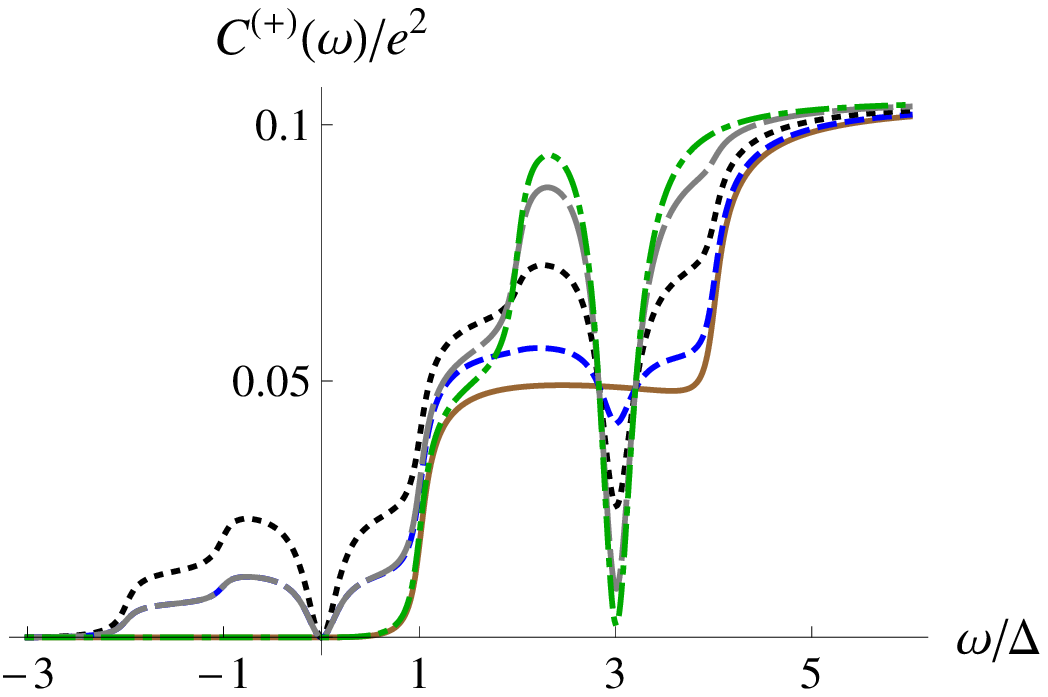} \\
 \includegraphics[width=8cm]{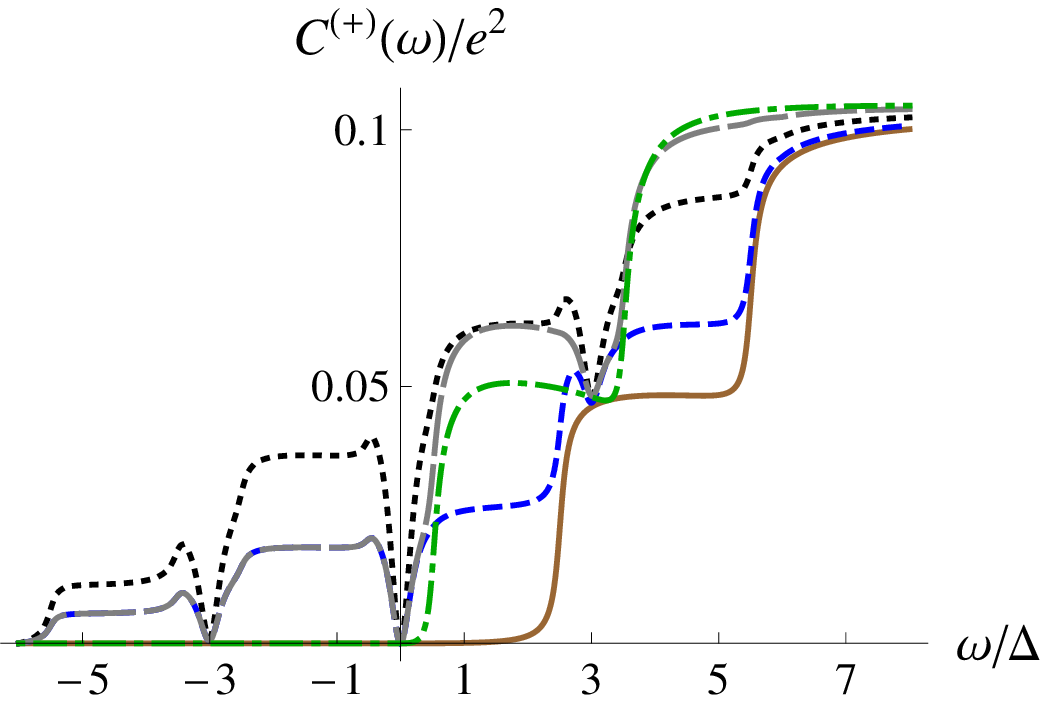} \\
 \includegraphics[width=8cm]{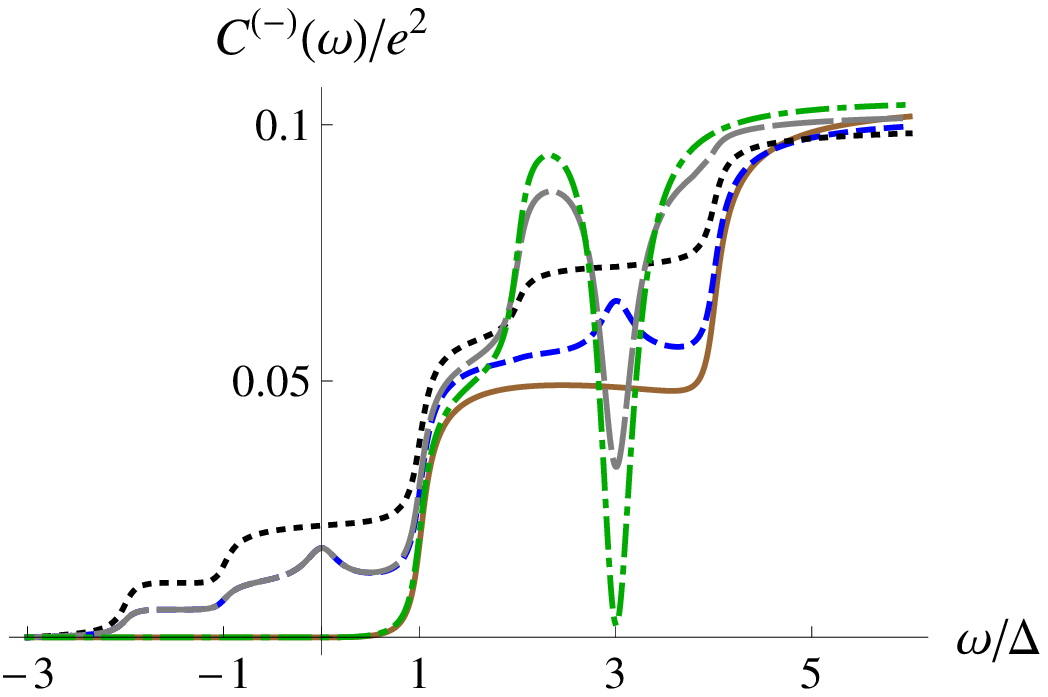}\\
 \includegraphics[width=8cm]{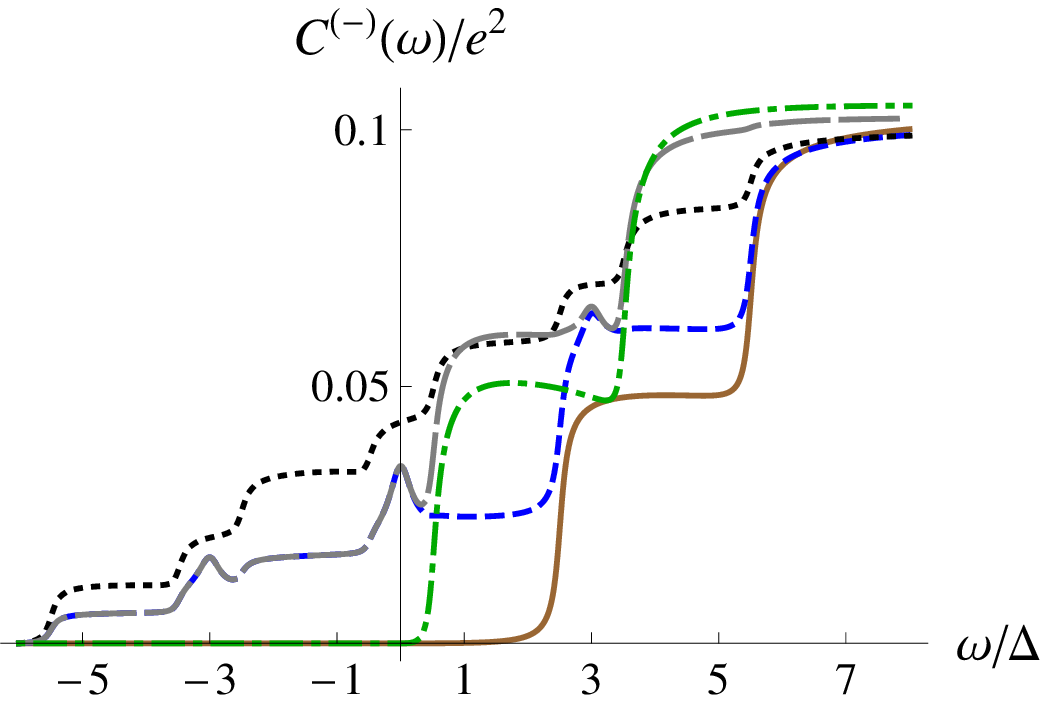}
    \caption{(color online) The noise spectra $C^{(+)}(\omega)$ (top figures) and  $C^{(-)}(\omega)$ (bottom figures) of the biased dot at zero temperature with $\epsilon_1 = \Delta$ and $\epsilon_2=-2\Delta$. The Fermi energy is set to $0.5$ in all figures.  The different figures are for for eV=3 (first and third figures from the top) and  6 (second and forth figures from the top). The five curves correspond to $\Gamma_L=0.1$ and $\Gamma_R=0$ (brown continuous curves), $\Gamma_L=0.0745$ and $\Gamma_R=0.0255$ (blue dashed curves), $\Gamma_L=\Gamma_R=0.05$ (black dotted curves), $\Gamma_L=0.0255$ and $\Gamma_R=0.0745$  (gray long dashed curves), and $\Gamma_L=0$ and $\Gamma_R=0.1$ (green  dotted-dashed curves). All energies are in units of $\Delta$.} \label{FigNoiseV}
\end{figure}

At finite biases Eq. \eqref{CMINT} is not applicable anymore and one needs to use
Eqs. \eqref{AA'}-\eqref{AA} to obtain the different contributions of
the inter- and intra-lead processes to the noise spectra.
As  pointed out  in
Sec. \ref{SCAT}, a finite bias modifies the integration bounds in
Eqs. \eqref{AA'}-\eqref{AA}, and as a result the  steps in the noise
spectra are not only shifted,
but split as well, as can be seen in
Fig. \ref{FigNoiseV}. The {\em dip} however, (which occurs roughly at
$\omega=|\epsilon_1-\epsilon_2|$) is not affected by the finite bias, in agreement with the findings of
Ref. ~\onlinecite{MARCOS}.
When only one of the levels lies in-between the  two chemical potentials of the leads
(this scenario corresponds to $eV=3$ for the parameters used in  Fig. \ref{FigNoiseV}) the dip in $C^{(+)}$
becomes sensitive to the ratio $\Gamma_L/\Gamma_R$, and the dip in $C^{(-)}$
can turn into a peak for certain values of $\Gamma_L/\Gamma_R$. When the bias
is large enough, so that the two levels are well between the two chemical potentials
of the leads (for the parameters used in Fig. \ref{FigNoiseV}, this occurs for $eV=6$), the dip in
$C^{(+)}$  no longer depends on the ratio $\Gamma_L/\Gamma_R$,
and the  dip in $C^{(-)}$ turns into a peak. Finally, when the bias is large
enough for the noise to become non zero at $\omega=-|\epsilon_1-\epsilon_2|$,
an additional dip [peak]  appears in $C^{(+)}$ [$C^{(-)}$] at negative frequencies.

The effect of both a finite bias voltage and a finite temperature
is exemplified in Fig.  \ref{FigNoiseTV}. Interestingly enough, while the steps in the spectra are smeared by the temperature, the
structure around
$\omega=|\epsilon_1-\epsilon_2|$ seems to be rather immune.
The dependence of this (dip) structure on the asymmetry of the dot
couplings to the leads may be therefore amenable to experimental
verification, making the noise spectra a possible tool to study the energy spectrum of quantum dots.

\begin{figure}
\includegraphics[width=8cm]{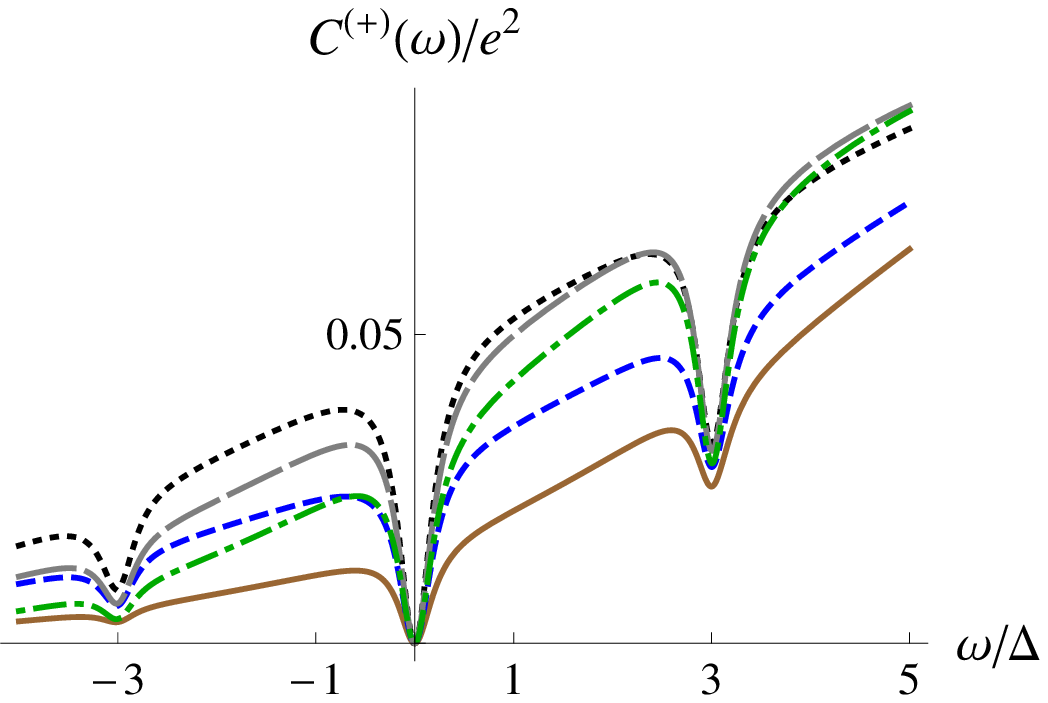} \\
\includegraphics[width=8cm]{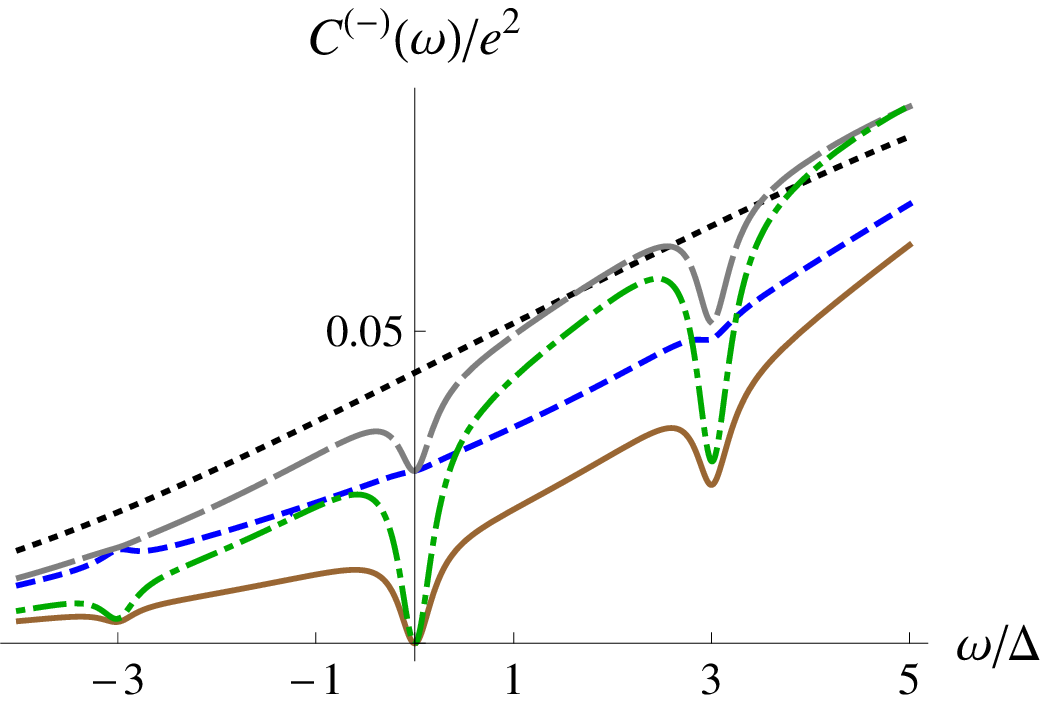}
 \caption{Same as  Fig. \ref{FigNoiseV},   with $T=1.5\Delta $ and $eV=6 \Delta$.} \label{FigNoiseTV}
\end{figure}

\begin{acknowledgments}
We thank Z. Ringel for helpful discussions. The work was supported by the German Federal
Ministry of Education and Research (BMBF) within the framework of
the German-Israeli project cooperation (DIP), and by  the US-Israel
Binational Science Foundation (BSF).
\end{acknowledgments}

\appendix{}
\section{A possible experimental observation of the  Fock terms}
\label{APPENDIX A}

As can be appreciated from Eq. (\ref{hdoteff}),
the Fock terms  couple together  the (Hartree-modified) single-electron
energies on the dot.   Thus, the original onsite levels  are modified by this coupling.
A possible way to detect   these terms might be to monitor the outcome of
such a coupling between the dot's levels. For  example,  one may ask how
the I-V characteristics are modified when this off-diagonal coupling is changed. This may
be related to the recent experiment  of
K\"{o}nemann {\it et al.},   \cite{KYNEMANN}
in which the slope of the steps in the differential conductance
depended on the polarity of the bias voltage.
These authors explain their results by relating them to the broadening of the levels (see below).

\begin{figure}[b]
\includegraphics[width=8cm]{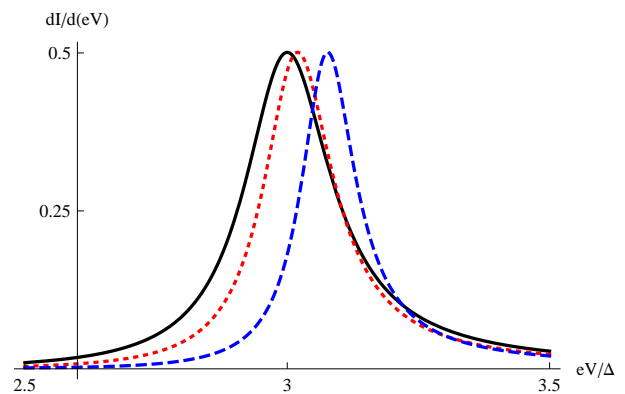}
\caption{(color online) The peak in the differential conductance as a function of the bias voltage, for various couplings $J$:  $J=0$ (black continuous curve), $J=0.1$ (red dotted curve), and $J=0.2$ (blue dashed curve). Parameters: $\epsilon_{\rm{F}}=0.5$, $\epsilon_1=1$,  $\epsilon_2=2$, $\Gamma_L=\Gamma_R=0.05$. (All energies are in units of the level spacing $\Delta$.)}
\label{FIGdIdV}
\end{figure}

In order to explore this possibility, we consider the simplest case of a (non interacting)
two-level dot and calculate
its differential conductance at zero temperature, in particular its dependence on the off-diagonal coupling of the levels.  By the Landauer formula, \cite{LANDAUER,IMRY} the differential conductance is
\begin{align}
\frac{dI}{d(eV)}=\frac{e}{4\pi}\Bigl ({\cal T}_{\rm TLD}^{}(\mu_{L}^{})+{\cal T}_{\rm TLD}^{}(\mu_{R}^{})\Bigr )\ ,
\end{align}
where ${\cal T}_{\rm TLD}$ is the transmission of the two-level dot. Denoting the two (original) onsite energies by $\epsilon_{1}$ and $\epsilon_{2}$, their coupling by $-J$, and assuming that they are identically coupled to each of  the two leads, the transmission is

\begin{align}
{\cal T}^{}_{\rm TLD}(E)=\Gamma^{}_{L}\Gamma^{}_{R} |g^r(E) |^{2}\ ,\label{TLD}
\end{align}
where $\Gamma=\Gamma_{L}+\Gamma_{R}$ and $g^r(E)$ is given in Eq. \eqref{COUPLED g}. (Note that we have assumed
here that the widths $\Gamma_{L}$ and $\Gamma_{R}$ are independent of the energy.) The differential conductance is plotted in Fig.
\ref{FIGdIdV}. As can be seen there,  increasing the coupling $J$ between the levels narrows down the peak of the differential conductance, in a rough agreement with the observations and interpretations of K\"{o}nemann {\it et al.}\cite{KYNEMANN}
Indeed, as  can be noted from Eqs. ({\ref{COUPLED g}) and (\ref{TLD}), the off-diagonal coupling $J$ modifies the broadening of the energy levels. In the simple case treated here, the width of the upper level (say $\epsilon_{1}$)  is reduced, from $\Gamma$ to $\Gamma (1-b)$, while the width of the lower level is enhanced, to be
$\Gamma (1+b)$, with $b=2J/\sqrt{4J^2+(\epsilon_1-\epsilon_2)^2}$. This observation helps us to understand the (small) differences between the curves in Figs.   \ref{AVER} and \ref{AVER1} calculated within the Hartree approximation alone, and those computed within the Hartree-Fock approximation. Consider for example the  diagonal averages, corresponding to the level occupancies. As is known, \cite{DONIACH} their variation with the interaction strength is determined by the levels' width; the Fock terms induce modifications in those widths and consequently change slightly the curves.


\begin{thebibliography}{999}







\bibitem{LANDAUER}
R. Landauer, Nature (London) \textbf{392}, 658 (1998).


\bibitem{IMRY}
Y. Imry, \emph{Introduction to Mesoscopic Physics}, 2nd ed. (Oxford University Press, Oxford, 2002).

\bibitem{BEENAKKER}
C. Beenakker and C. Sch\"{o}nenberger, Phys. Today {\bf56}, 37 (2003).

\bibitem{BLANTER}
Ya. M. Blanter and M. B\"{u}ttiker, Phys. Rep. {\bf336}, 1 (2000).

\bibitem{AVERIN}
D. V. Averin and J. P. Pekola, Phys. Rev. Lett. {\bf104}, 220601 (2010).


\bibitem{CLERK}
A. A. Clerk, M. H. Devoret, S. M. Girvin, F. Marquardt, and R. J. Schoelkopf, Rev. Mod. Phys. {\bf82}, 1155 (2010).
\bibitem{GAVISH}

U. Gavish, Y. Levinson, and Y. Imry, Phys. Rev. Lett. {\bf 87},
216807 (2001); U. Gavish, Y. Levinson, and Y. Imry, Phys. Rev. B {\bf 62},
R10637 (2000).

\bibitem{AGUADO}
R. Aguado and L. P. Kouwenhoven, Phys. Rev. Lett. {\bf84}, 1986 (2000).





\bibitem{DEBLOCK}
R. Deblock, E. Onac, L. Gurevich, and L. Kouwenhoven, Science \textbf{301}, 203 (2003).

\bibitem{GUSTAVSSON}
S. Gustavsson, M. Studer, R. Leturcq, T. Inn, K. Ensslin, D. C. Driscoll, and A. C. Gossard, Phys. Rev. Lett. \textbf{99}, 206804 (2007).

\bibitem{DEBLOCK 2}
J. Basset, H. Bouchiat, and R. Deblock, Phys. Rev. Lett. {\bf105},
166801 (2010) .

\bibitem{STEPS AND DIPS}
O. Entin-Wohlman, Y. Imry, S. A. Gurvitz, and A. Aharony, Phys. Rev. B {\bf75}, 193308 (2007).

\bibitem{ME}

E. A. Rothstein, O. Entin-Wohlman, and A. Aharony, Phys. Rev. B
{\bf79}, 075307 (2009).

\bibitem{DONG}
B. Dong, X. L. Lei, and N. J. M. Horing, J. Appl. Phys. {\bf104}, 033532 (2008).

\bibitem{WU}
B. H. Wu and C. Timm, Phys. Rev. B {\bf81}, 075309 (2010).

\bibitem{MARCOS}
D. Marcos, C. Emary, T. Brandes, and R. Aguado, Phys. Rev. B {\bf83}, 125426 (2011).

\bibitem{HAMMER}
J. Hammer and W. Belzig, Phys. Rev. B {\bf 84}, 085419 (2011).

\bibitem{THIELMANN}
A. Thielmann, M. H. Hettler, J. K\"{o}nig, and G. Sch\"{o}n, Phys. Rev. B {\bf68}, 115105 (2003).


\bibitem{GUSTAVSSON2}
S. Gustavsson, R. Leturcq, B. Simovi\v{c}, R. Schleser, T. Ihn, P.
Studerus, and K. Ensslin, Phys. Rev. Lett. {\bf 96}, 076605
(2006).
\bibitem{KYNEMANN}
J. K\"{o}nemann, B. Kubala, J. K\"{o}nig, and R. J.
Haug, Phys. Rev. B \textbf{73}, 033313 (2006).

\bibitem{ZARCHIN}
O. Zarchin, M. Zaffalon, M. Heiblum, D. Mahalu, and V. Umansky, Phys. Rev. B {\bf 77}, 241303(R) (2008).

\bibitem{MOCA}
C. P. Moca, P. Simon, C. H. Chung, and G. Zar\'{a}nd, Phys. Rev. B {\bf 83}, 201303 (2011).

\bibitem{Alhassid}
Y. Alhassid, H. A. Weidenm\"{u}ller, and A. Wobst, Phys. Rev. B
{\bf 72}, 045318 (2005).

\bibitem{Kunz}  H. Kunz and R. Rueedi, Phys. Rev. A {\bf 81},
032122 (2010).

\bibitem{SINDEL} M. Sindel, A. Silva, Y. Oreg, and J. von
Delft, Phys. Rev. B {\bf 72}, 125316 (2005).

\bibitem{GOLDSTEIN}
M. Goldstein and R. Berkovits, New J. Phys. {\bf9}, 118 (2007).

\bibitem{CATELANI}
G. Catelani and M.G. Vavilov, Phys. Rev. {\bf B} 76, 201303(R) (2007).

\bibitem{GABELLI}
J. Gabelli, G. Feve, J.-M. Berrior, B. Placais, A. Cavanna, B.
Etienne, Y. Jin, D. C. Glattli, Science \textbf{313}, 499 (2006).

\bibitem{BUTTIKER3}
M. B\"uttiker and S. E. Nigg, Nanotechnology, \textbf{18}, 044029 (2007)

\bibitem{ZOHAR}
Z. Ringel, Y. Imry, and O. Entin-Wohlman, Phys. Rev. B \textbf{78}, 165304 (2008).



\bibitem{BLUNTER}
Ya. M. Blanter and M. B\"{u}ttiker, Phys. Rep. {\bf336}, 1 (2000).

\bibitem{BUTTIKER1}
M. B\"{u}ttiker, Phys. Rev. B \textbf{46}, 12485 (1992).

\bibitem{BUTTIKER2}
M. B\"{u}ttiker, A. Pr\^{e}tre, and H. Thomas, Phys. Rev. Lett. \textbf{70}, 4114 (1993).

\bibitem{MATVEEV}

K. A. Matveev, Phys. Rev. B \textbf{51}, 1743 (1995).


\bibitem{BRUUS}

H. Bruus and K. Flensberg, \emph{Many-Body Quantum Theory in Condensed Matter Physics}, 1st ed. (Oxford University press, New York, 2004).


\bibitem{BEENAKKER2}
P. W. Brouwer and C. W. J. Beenakker, Phys. Rev. B \textbf{55}, 4695 (1997).

\bibitem{BUTTIKER4}
M. B\"uttiker, Phys. Rev. B \textbf{41}, 7906 (1990).


\bibitem{DONIACH}
S. Doniach and E. H. Shondheimer, \emph{Green's functions for solid state physicists}, 3rd ed. (Imperial College Press, London, 1998).


\end{thebibliography}
\end{document}